\documentclass[11pt]{article}

\usepackage[preprint]{acl}
\setcitestyle{numbers,square,comma}
\hypersetup{pdftitle={Auditing LLM Alpha-Factor Mining: Metric Provenance and Full-Cohort Reanalysis},pdfauthor={Yufeng He},pdfsubject={AI4F 2026 workshop submission}}

\usepackage[T1]{fontenc}
\usepackage[utf8]{inputenc}
\usepackage{mathptmx}
\usepackage[scaled=0.92]{helvet}
\usepackage{courier}

\usepackage{booktabs}
\usepackage{graphicx}
\usepackage{amsmath}
\usepackage{microtype}
\usepackage{inconsolata}
\usepackage{enumitem}

\title{Auditing LLM Alpha-Factor Mining: \\ Metric Provenance and Full-Cohort Reanalysis}

\author{Yufeng He \\ The University of Hong Kong \\
  \texttt{he-yufeng@connect.hku.hk}}

\begin{document}
\maketitle
\begin{abstract}
Reliable evaluation of LLM-based alpha mining requires more than positive backtest summaries. We audit a preserved project at three levels: historical run counters, outcome retention, and a separately scoped public reanalysis. Three historical runs report 67, 20, and 11 keep decisions, but the comparison mixes metric populations: a mean over kept candidates is compared with a mean of per-round maxima, and successful evaluations with all returned outcomes. These artifacts cannot identify an architecture advantage. A released 1,837-expression catalog also excludes non-positive OOS outcomes, making its zero-reversal rate non-informative. We then account for all 1,047 unique expressions in a frozen, IS-screened preserved pool using forty A-shares and 2024--2025 rank associations. A versioned public-operator correction yields 923 scorable records, 99 with insufficient valid dates, and 25 unsupported, without modifying the other 990 prior diagnostics. Among scorable records, 671 have positive and 252 negative signed mean rank-IC. Simple untrained references also show positive associations, so this distribution does not establish incremental agent skill or profitability. The case study motivates explicit metric grains, immutable cohort accounting, and provenance-aware interpretation rather than architectural rankings from incomplete records.
\end{abstract}

\section{Introduction}
LLM-based factor mining combines expression generation, a backtester, and selection rules. Its appeal is practical: models can propose candidate signals, use feedback, and revise a search direction. Yet a pipeline can produce plausible summary tables without preserving the evidence needed to interpret them. A field named \emph{Sharpe} does not specify which population was averaged; a count named \emph{backtested} does not establish whether failures are included; and a positive-only artifact cannot estimate the frequency of negative outcomes.

We study these problems through a preserved project containing a three-role planning--generation--review pipeline and a twenty-specialist evolutionary workflow. The historical reports initially suggested an architectural difference. Our audit instead asks: \emph{which conclusions do the counters, source paths, and retained artifacts support?} We do not introduce a new mining algorithm or claim superiority over prior systems.

The study contributes (1) a source-and-report audit exposing incompatible metric grains and incomplete configuration/state provenance (\S\ref{sec:historical}); (2) an outcome-retention audit of a 1,837-expression catalog (\S\ref{sec:oos}); and (3) a full-cohort public reanalysis of 1,047 frozen expressions, retaining non-scorable records and a versioned implementation correction (\S\ref{sec:public}). These are related project artifacts, not interchangeable experimental populations.

\section{Related Work}
Agentic generate--evaluate systems and evolutionary expression search are active approaches to alpha mining~\cite{alphaagent2025,quantaalpha2026,cognitivealpha2025}. They motivate careful evaluation of search outcomes, but our contribution is an artifact-level case study rather than a new search method or a benchmark against those systems. The findings do not imply that other systems lack adequate controls. They identify concrete ways in which this project's reporting and retained artifacts fail to support its initial interpretation.

\section{Workflow and Evidence Scope}
The preserved three-role workflow has a Researcher that plans a direction, an Implementor that generates and tests expressions, and a Reviewer. In the inspected source, keep/discard decisions are programmatic. The reviewer then attempts an LLM-generated next-step suggestion, with a rule fallback on failure. A role name or model configuration is not a verified successful model request, and we have not recovered the exact historical request graph.

The swarm has twenty specialists across twelve configured categories, an orchestrator, adaptive allocation weights, sign-flipped candidates, and periodic prompt-revision proposals. Both workflows use a functional price/volume DSL and a proprietary evaluator. The historical narrative lists common thresholds of Sharpe $\geq2$, Fitness $\geq0$, and Turnover $\leq500$. However, preserved source paths also have workflow-specific IC, complexity, pool-score, or correlation checks. Exact executed policy versions and matching initial pool states are not bound to the reports, so identical effective selection policies are not established.

The evidence consists of March 2026 run reports, a native OOS checkpoint, an outcome-filtered catalog, and a separately frozen list selected from 3,114 preserved input rows. Current source explains report construction but is not automatically treated as the exact historically executed commit. We corroborate its implications by recomputing statistics from printed report rows where possible.

\section{Historical Run-Report Audit}\label{sec:historical}
\subsection{Recoverable counters}
Each report contains twenty rounds. Table~\ref{tab:runs} preserves their recoverable counters without presenting unlike stages as one metric. Keep decisions are not proved counts of distinct factors surviving in the final pool. Missing cells are unavailable at a harmonized stage, not zero values.

\begin{table}[t]\centering\small
\caption{Recovered historical counters. P is the pipeline; S-A and S-B are two swarm runs. Different accounting stages occupy different rows.}\label{tab:runs}
\begin{tabular}{@{}lrrr@{}}
\toprule
Reported counter & P & S-A & S-B \\
\midrule
Rounds & 20 & 20 & 20 \\
Keep decisions & 67 & 20 & 11 \\
Pipeline tested candidates & 289 & -- & -- \\
Pipeline successful outcomes & 234 & -- & -- \\
Swarm generated proposals & -- & 887 & 876 \\
Swarm returned outcomes & -- & 816 & 461 \\
Keep decisions / round & 3.35 & 1.00 & 0.55 \\
\bottomrule
\end{tabular}
\end{table}

The contemporaneous narrative labels the pipeline generator as Claude Opus 4.6, S-A specialists as Claude Sonnet 4.5, and S-B specialists as Claude Opus 4.6. Some report labels are hard-coded in the template rather than recovered from requests. We retain them only as reported configuration labels, not a verified model-quality ranking or backend identity.

\subsection{Metric populations do not align}
The earlier mean-Sharpe comparison is not a common-population mean. The pipeline collector appends kept-candidate Sharpes, while the swarm collector appends positive per-round maxima. Three-decimal report entries reproduce averages of approximately 2.227 over 67 kept rows and 2.433 over 20 round maxima, matching the earlier rounded 2.23 versus 2.43 display. We do not retain this as a performance comparison.

The maximum also changes population: the pipeline's kept-candidate maximum is 2.449, whereas its evaluated round summaries reach 3.171; the swarm's 4.206 is an evaluated round maximum and need not be kept. These values cannot establish superior outlier discovery.

Likewise, 234 is the pipeline's reported successful-outcome count, but 816 is the swarm's returned-outcome count, including possible failures and additional sign-flip evaluations. Six S-A rounds have more outcome records than newly generated proposals. The thin report does not recover a success-only denominator, so $67/234$ versus $20/816$ is not a verified common-denominator acceptance-rate comparison.

\subsection{Categories and state provenance}
The pipeline's four category labels cover all round directives; the swarm's two productive labels cover mapped specialists with keep decisions. These are different populations. S-A's agent leaderboard attributes 8 keep decisions while the run counter is 20; S-B attributes 9 while its counter is 11. The source's sign-flip route can fall outside configured-specialist attribution, but the missing per-event attribution has not been reconstructed.

The comparison code runs the pipeline before the swarm without an explicit initial-pool snapshot/reset binding. Together with missing exact executed versions and request-level provenance, this prevents treating the reports as a controlled experiment. More seeds alone would not repair incompatible counters: future work must align accounting stages, selection policies, initial state and resource budgets before using independent runs for inference.

\section{Outcome-Retention Audit}\label{sec:oos}
The original analysis described 1{,}837 records as an OOS validation set. A read-only audit of the preserved artifacts and runner shows that this interpretation is not valid: the runner uses a looser preliminary screen and explicitly drops records with OOS Sharpe $\leq 0$ before writing the retained file. Moreover, only 1{,}037 rows in that file satisfy the manuscript's stated IS rule ($\textrm{Sharpe}\geq 2$, $\textrm{Fitness}\geq 0$, $\textrm{Turnover}\leq 500$).
\begin{table}[t]\centering\small
\caption{Read-only audit of OOS artifacts. The retained catalog was selected using OOS outcomes.}\label{tab:oos}
\begin{tabular}{lr}
\toprule
Artifact count & Value \\
\midrule
Checkpoint expressions & 2{,}016 \\
Negative-OOS checkpoint rows & 368 \\
Retained catalog expressions & 1{,}837 \\
Non-positive-OOS catalog rows & 0 \\
Checkpoint-only expressions & 368 \\
Checkpoint-only negative OOS & 368 \\
Catalog rows passing stated IS rule & 1{,}037 \\
\bottomrule
\end{tabular}
\end{table}
The checkpoint contains 2{,}016 unique expressions, including 368 with negative OOS Sharpe. All 368 expressions found only in the checkpoint are negative OOS; all 189 expressions found only in the retained catalog are positive OOS; and the 1{,}648 common expressions have no sign disagreement. The catalog's zero sign-reversal rate and its decay summaries are therefore properties of outcome-based retention, not evidence that the mined factors generalize.

A clean OOS evaluation remains pending. It must recover and hash the complete pre-OOS pool, pre-specify an IS-only screen, perform any deduplication using IS data, freeze the expression list before OOS access, checkpoint every response, and preserve every numerical OOS outcome and error regardless of sign. Numerical OOS and generalization claims should not be restored until that run exists.

\section{Full-Cohort Public Reanalysis}\label{sec:public}
We separately evaluate a frozen list of 1,047 unique expressions selected from 3,114 preserved input rows using the stated IS screen. This does not establish a census of all generated candidates or a historically blind holdout. The broader preserved pool is not a matched replay of the three historical keep totals. All IDs are retained regardless of subsequent public-evaluator outcome; no expression is chosen by its new score.

The new snapshot contains the same ordered 40-stock convenience universe over 727 trading dates in 2023--2025. We acquire both unadjusted and provider-adjusted daily views through BaoStock 0.9.3 and verify all 80 files, their byte hashes, date/stock/adjustment identities, and shared quantity and trading-status fields. Features use nominal prices, shares, and nominal traded amount; the adjusted-price view is reserved for labels. Two non-trading stock-day cells are masked. We neither forward-fill missing observations nor substitute a smaller acquired stock prefix.

For each 2024--2025 date $t$, the label is the provider-adjusted close change over five trading-calendar rows,
\[
y_{i,t}=\frac{P^{\mathrm{adj}}_{i,t+5}}{P^{\mathrm{adj}}_{i,t}}-1.
\]
Both endpoint prices must be finite and positive and both endpoints tradable. This is not an investor total-return or executable-P\&L claim; the last five captured dates have no label. Daily rank-IC is the Pearson correlation of average-tie cross-sectional ranks. We require at least five eligible stocks, three distinct signal values, and two distinct labels. The three-signal-value threshold is a chosen diagnostic convention, not a claim that binary signals cannot have a mathematically defined correlation. A factor is scorable only with at least 20 valid dates. Its mean/SD ratio, if reported, is unannualized and is not a trading Sharpe.

The frozen public implementation initially produces 57 numerical errors: its conditional operator attempts to interpret a panel as one Boolean when the true branch is scalar. We preserve this v1 run and add a versioned scalar-broadcasting correction with the existing elementwise conditional semantics. All 1,047 records are then re-evaluated, rather than rerunning only promising failures. The other 990 diagnostic records remain unchanged; 56 corrected records become scorable and one remains insufficiently observed.

\begin{table}[t]\centering\small
\caption{Full frozen cohort under the corrected public evaluator. Non-scorable records remain in the denominator; no missing result is assigned zero alpha.}\label{tab:public}
\begin{tabular}{lr}
\toprule
Outcome & Records \\
\midrule
Scorable signed rank-IC & 923 \\
Insufficient valid dates & 99 \\
Unsupported operator or variable & 25 \\
\textbf{All frozen records} & \textbf{1,047} \\
\midrule
Positive mean rank-IC (of 923) & 671 \\
Negative mean rank-IC (of 923) & 252 \\
Median mean rank-IC (of 923) & 0.0148 \\
\bottomrule
\end{tabular}
\end{table}

Table~\ref{tab:public} is descriptive. Of the 923 scorable expressions, 835 have all 480 label-available target dates; the other 88 have varying coverage, which further limits cross-factor comparisons. We verify file identities and nine fixed record/date rank-IC cells with a separate tie-ranking/Pearson calculation. Factors and overlapping five-day labels are not independent observations, so we do not attach an independent-factor significance test to the aggregate distribution. These results are not native OOS Sharpe, are not directly interchangeable with the outcome-filtered catalog's metrics, and do not identify an architecture effect.

\subsection{Fixed reference-signal context}
After the cohort analysis, we specify six fixed reference signals before computing their outcomes. All are reported in Table~\ref{tab:references}; signs and horizons are not selected based on these scores. They use the same snapshot and scoring convention. Reversal references negate nominal trailing close changes; the twenty-day reference retains the positive orientation.

\begin{table}[t]\centering\small
\caption{Post-reanalysis descriptive references, each with 480 valid dates. These untrained signals are context, not a matched-sample ranking against the factor cohort.}\label{tab:references}
\begin{tabular}{@{}lr@{}}
\toprule
Fixed reference & Mean signed rank-IC \\
\midrule
Nominal close level & $-0.0218$ \\
Share volume & $0.0234$ \\
Nominal traded amount & $0.0021$ \\
One-day nominal reversal & $-0.0007$ \\
Five-day nominal reversal & $-0.0155$ \\
Twenty-day nominal change & $0.0229$ \\
\bottomrule
\end{tabular}
\end{table}

Volume and recent-change references also have positive associations. Therefore a positive IC alone does not establish incremental agent skill; we have not estimated incremental prediction beyond these references. Factors can have different eligible dates and stocks, so their cohort median and the standalone reference means do not establish matched-sample outperformance. No reference is selected for a profitability or significance claim.

\section{Implications and Limitations}
The defensible contribution is an evaluation-audit case study, not an architectural ranking. Historical reports preserve different keep counters but not verified common-grain Sharpe, rate, or category comparisons. Single runs, unequal budgets, unbound starting state and uncertain executed configurations further limit interpretation. The public cohort comes from a broader preserved pool and cannot be assigned back to the three run conditions from the available records.

The reanalysis completes data collection and assigns an explicit outcome to every frozen expression. It still covers one market and a convenience stock universe. It uses public rather than native semantics; factor coverage varies and five-day labels overlap. The original proprietary evaluator and full historical request traces are not released. The reference checks were planned after the cohort analysis, not as part of its original primary protocol. A stronger study would fix common masks and comparators, retain generation and run identities, and test repeated systems under matched budgets and a suitable temporal design.

\section{Conclusion}
Preserved artifacts can support an audit even when they cannot support a performance ranking. This case reveals metric-population mismatches, outcome-conditioned retention, and a public conditional-operator defect. A complete, versioned reanalysis retains all 1,047 IDs, including 124 non-scorable records, without treating positive associations as profitability or architecture effects. Reliable future comparisons require aligned definitions and provenance before additional scale or significance testing.

\section*{Ethics Statement}
The new analysis uses publicly accessible historical market data for research, makes no live-trading or financial-advice claim, and sends no trading orders. Failed and non-scorable records are preserved. Proprietary components and raw provider data are not represented as publicly redistributable artifacts; the SDK license is not a data-redistribution grant. Model labels are reported with provenance limits rather than as verified backend identities.

\paragraph{AI-assistant disclosure.} AI assistants supported literature review, audit design, implementation, statistical checking and manuscript preparation. Numerical findings are computed from preserved source records and versioned reanalysis outputs, not generated as substitutes for observations. No human-annotation study or independent human review is claimed. The author remains responsible for the content.

\end{document}